# On the nature of Surface States Stark Effect at clean GaN(0001) surface

Paweł Kempisty[1], and Stanisław Krukowski[1,2],

[1]*Institute of High Pressure Physics, Polish Academy of Sciences, Sokołowska 29/37, 01-142 Warsaw, Poland*

[2]*Interdisciplinary Centre for Mathematical and Computational Modelling, University of Warsaw, Pawińskiego 5a, 02-106 Warsaw, Poland*

**Abstract**

Recently developed model allows for simulations of electric field influence on the surface states. The results of slab simulations show considerable change of the energy of quantum states in the electric field, i.e. Stark Effect associated with the surface (SSSE – Surface States Stark Effect). Detailed studies of the GaN slabs demonstrate spatial variation of the conduction and valence band energy revealing real nature of SSSE phenomenon. It is shown that long range variation of the electric potential is in accordance with the change of the energy of the conduction and valence bands. However, at short distances from GaN(0001) surface, the valence band follows the potential change while the conduction states energy is increased due to quantum overlap repulsion by surface states. It is also shown that at clean GaN(0001) surface Fermi level is pinned at about 0.34 eV below the long range projection of the conduction band bottom and varies with the field by about 0.31 eV due to electron filling of the surface states.

# I. Introduction

Investigations of physical properties of Ga-terminated GaN(0001) surface are interesting from the point of basic research and also for possible applications. The properties of GaN(0001) surface could be possibly strongly affected by its preparation therefore surface preparation techniques were discussed in detail by Bermudez et al. [1]. Accordingly, measurements of band bending ($e_oV_s$) at GaN (0001) surface brought plethora of different results. First data, reported by Dhesi et al.[2] and Valla et al.,[3] obtained in the same laboratory were: $e_oV_s$ = 2.1 ±0.1 eV and close to 1.5 eV, respectively. Such difference indicate on the difficulty on preparation of ideally clean surface. In the meantime Wu et al. obtained band bending $e_oV_s$ = ±0.75 eV, upward and downward for p-GaN and n-GaN, respectively.[4] Later on Long and Bermudez confirmed results of Wu et al for n-type GaN for which Fermi level is pinned by the surface state at 2.55 eV above Valence Band Maximum (VBM).[5] The difference for p-GaN was probably caused by technical difficulties in measurement of downward band bending by UPS technique. In the same year, the two other different values of band bending were published: for n-type GaN(0001) surface by Kočan et al. reporting $e_oV_s$ = -0.46 eV,[6] and by Plucinski et al. showing no band bending at all.[7] In the following years two more sets of data were reported by Cho et al. $e_oV_s$ = 1.0 eV[8] and by Widstrand et al. where essentially no band bending (flat bands) was obtained again.[9] Therefore direct measurement values show considerable scattering of data which need to be confronted with other estimates. Such results were obtained from contact electroreflectance (CE) studies from which it was deduced that at GaN(0001) surface in air Fermi level was pinned 0.55±0.05 eV below the conduction band minimum (CBM).[10] Unfortunately, no data on the Fermi level at vacuum have been obtained so far.

The electronic properties of the clean GaN(0001) surface include dispersion of the surface states that were determined by angle resolved photoelectron spectroscopy (ARPES) detecting the two bands of surface states. The first, of negligible dispersion, is located at about 1.5 eV below VBM. [7, 9, 11] The second, highly dispersive band, also located below VBM, has dispersion over 4 eV[11]. These bands are removed by exposition of the surface to activated hydrogen or oxygen thus the weekly dispersive band is composed of the gallium dangling bonds.[2]

As these dispersion relations are intimately related to the surface structure, Quantum Mechanical Density Functional Theory (QM DFT) calculations were used in investigations of the properties of clean GaN(0001) surface [12,13,14,15]. Initially ab initio investigations showed the bare GaN(0001) surface is terminated by triply bonded Ga atoms without any reconstruction.[12,13] Initially these findings were confirmed also by our calculations in which the dependence on the electric field was accounted for[16]. Recently it was found however, that the 2 x 1 row and valley structure is more stable having the total energy slightly lower than the unreconstructed surface.[17] The reconstruction was identified to be due to change of the bonding to $sp^2$ hybridization of quantum surface bonding states, lowering their energy due to higher contribution of s orbitals and leaving the remaining p orbital empty. Nevertheless the measurement of the dispersion-free surface states close to VBM was not confirmed.

It is therefore important to determine the relation between the electric field, the existence of charged surface states and structural properties of the GaN(0001) surfaces. The most reliable DFT calculations provided essentially the same results,[12,13] the surface states is characterized by large dispersion close to 2 eV, with the minimum at Γ point. The Fermi level is pinned by the surface states. It has to be added however, that these calculations did not account the influence of the fields induced by charged surface states. Recently we have developed new model allowing to simulate the electric field at the surface in the slab model.

[16,17] By proper manipulation of the H-termination atoms the field at the surface could be changed with the appropriate change of the electric charge at the surface states. It was shown also that the position of the surface states and the band volume states is changed by the change of this field. Aftermath, the same model was applied to the influence of hydrogen adsorption at the GaN(0001) surface on the dispersion relation. Similarly, the properties of the SiC surface was modeled discerning the change of the relative energy of the volume and surface states within the slab model, the phenomenon which was denoted as Surface States Stark Effect (SSSE). [18] The model was not widely used which is disappointing as it could potentially bring deeper insight into the electronic properties of semiconductor surfaces. A possible obstacle may be related to relatively terse description of the model, and the underlying physics. Present paper is intended to amend this by detailed discussion of the electric and quantum properties of clean GaN(0001) surface described within this model.

## II. Simulation procedure

In all calculations reported below a freely accessible DFT code SIESTA, combining norm conserving pseudopotentials with the local basis functions, was employed[19][20][21]. The basis functions in SIESTA are numeric atomic orbitals, having finite size support which is determined by the user. The pseudopotentials for Ga, H and N atoms were generated, using ATOM program for all-electron calculations. SIESTA employs the norm-conserving Troullier-Martins pseudopotential, in the Kleinmann-Bylander factorized form.[22][23] Gallium 3d electrons were included in the valence electron set in explicit manner. The following atomic basis sets were used in GGA calculations: Ga (bulk) - 4s: DZ (double zeta), 4p: DZ, 3d: SZ (single zeta), 4d: SZ; Ga (surface)- 4s: DZ, 4p: DZ, 3d: SZ, 4d: SZ, 5s: SZ; N (bulk) - 2s: TZ (triple zeta), 2p: DZ;  N (surface)- 2s: TZ, 2p: DZ, 3d: SZ, 3s: SZ; H - 1s: QZ (quadruple zeta), 2p: SZ and H (termination atoms)  1s: TZ, 2p: SZ. The following values for the lattice constants of bulk GaN were obtained in GGA-WC calculations (as exchange-correlation functional Wu-Cohen (WC) modification of Perdew, Burke and Ernzerhof (PBE) functional [24] [25] was used): a = b = 3.2021 Å , c = 5.2124 Å. These values are in good agreement with the experimental data for GaN: a = 3.189 Å and c = 5.185 Å. [26] All presented dispersion relations are plotted as obtained from DFT calculations, hence in order to obtain the quantitative agreement with experimentally measured values, all calculated DFT energies, should be rescaled by approximate factor $\alpha = E_{g-exp}/E_{g-DFT}$=3.4eV/2.13eV ≈ 5/3 ≈ 1.6. In order to obtain the Hamiltonian matrix elements, we employed a grid in real space, which was obtained using a mesh cutoff of 275 Ry. Integrals in k-space were performed using 3x3x1 Monkhorst-Pack grid for slab with lateral size 2x2 unit cell and only Γ-point for larger slabs. [27] As a convergence criterion, terminating SCF loop, the maximum difference between the output and the input of each element of the density matrix was employed being equal or smaller than $10^{-4}$. Relaxation of atomic position is stopped when the forces on the atoms become smaller than 0.04 eV/Å.

In application of the SIESTA the efficient solver of Poisson equation, based on FFT method, implemented by SIESTA authors, was used. The method is based on assumption of the PBC at all sides of the simulated volume. Laplace correction method was employed, which improves convergence of SCF loop alleviating this condition[17]. This method is considerably better than the dipole correction[28], especially for thick Ga-N slabs.

### III. Electric fields, density of states, and surface states .

In most cases semiconductor surfaces are occupied by excessive charge leading to upward and downward band bending. Frequently, these fields extends over the tens or hundreds of atomic layers, therefore direct ab initio simulation of the charged layers at surfaces are impossible.

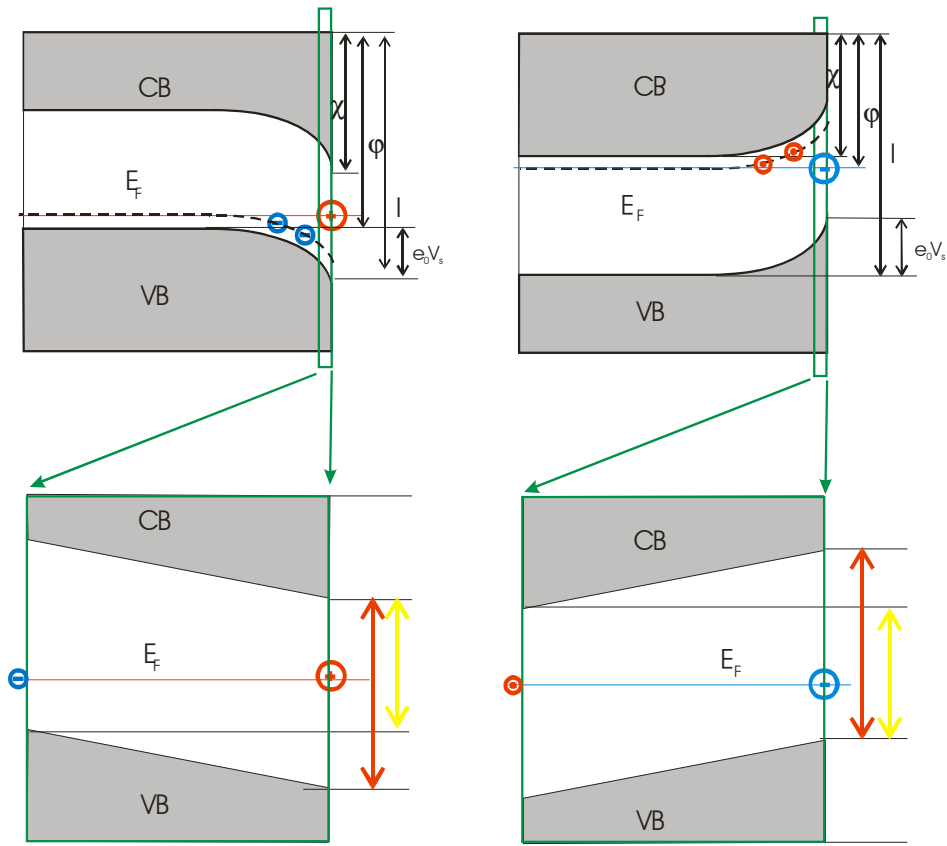

Fig.1. Top diagrams: band arrangement at the surface in p-type GaN (left) and n-type GaN (right). Bottom – representation of the fields and the location of the charges at the slab terminations: left – termination surface, right – real surface. By right arrow the bandgap is denoted, by yellow arrow – the bandgap reduced due to projection of electric field skewed band states.

Fortunately, new method of simulation of the electric fields was devised recently which, as shown in Fig. 1, relies on a representation of thin subsurface electric layer in the slab, mimicking the outside field by appropriate conditions at the opposite termination surface. As in the case of parallel plate capacitor, the field does not depend on the distance between the plates, so electric conditions in the narrow layer at the surface could be simulated exactly.

Thus a natural condition of successful representation of the surface electric properties is linear dependence of the average electric potential within the slab. The charge at the termination side could be changed by use of different charge of hydrogen termination atoms. The three typical distributions of the electric potential, obtained for 10 and 20 Ga-N double atomic layers (DALs) slab, are presented in Fig. 2.

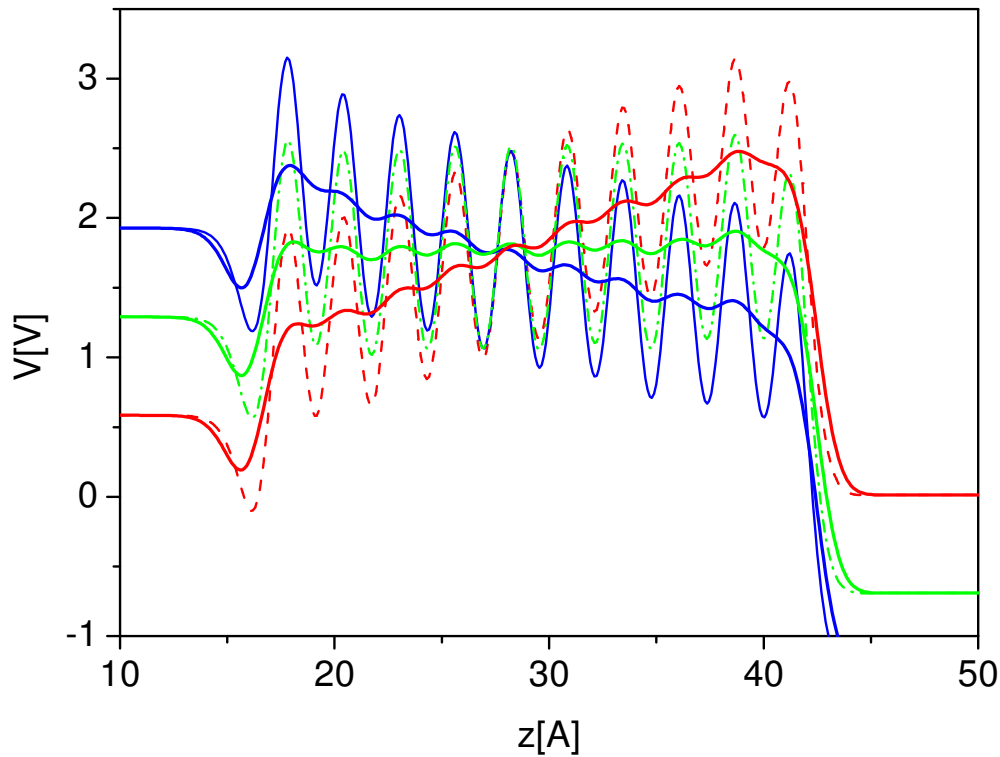

Fig.2. Potential distribution across across 20 Ga-N DALs GaN slab terminated by fractional charge hydrogen atoms located at d = 1.07 Å distance from the bottom N atoms: red lines Z = 0.70 e, green lines Z = 0.73 e and blue lines Z = 0.78 e. Thin lines – denote the potential along the channeling path, solid thick lines – averaged over $\Delta z$ = 2.7 Å.

As it is shown in Fig. 2, a supposed linear potential distribution is obtained in some cases while in the others, as shown below for the bigger slab, the average profile is strongly nonlinear , i.e. the field is not constant within the slab. A close inspection shows also that for 10 DAL slab, the average potential for the case Z = 0.78e deviates from strictly linear dependence. Naturally, any linear potential distribution could be obtained only in the case in which net electric charge is not accumulated in the slab interior. Since ideal crystal structure is used in these DFT slab simulations, no charge could be associated with the defects therefore the charge can originate either from a nonuniform polarization or from occupation of the band states. Since the polarization contribution is negligible, the occupation due to Fermi level penetration into the valence/conduction band is the only factor that may contribute to nonlinear distribution of the potential.

As shown in Fig. 1, the effective bandgap is possibly reduced by projection of the quantum states across the slab, so that the positions of VBM for p- and CBM for n-GaN are misrepresented in band diagrams, such as shown in Fig.2. The spatial variation of the band states energy leads not only to the bandgap narrowing, but also to penetration of the Fermi level into the bands and occupation of the portion of the band states. This is shown in diagrams in Fig. 3 where the projected density of states and the band profiles are plotted. The projected density of states, is obtained by projection of the slab quantum states on the atomic wavefunctions in the consecutive layers. Positioning spatial location of these atoms reveals spatial variation of the band energy.

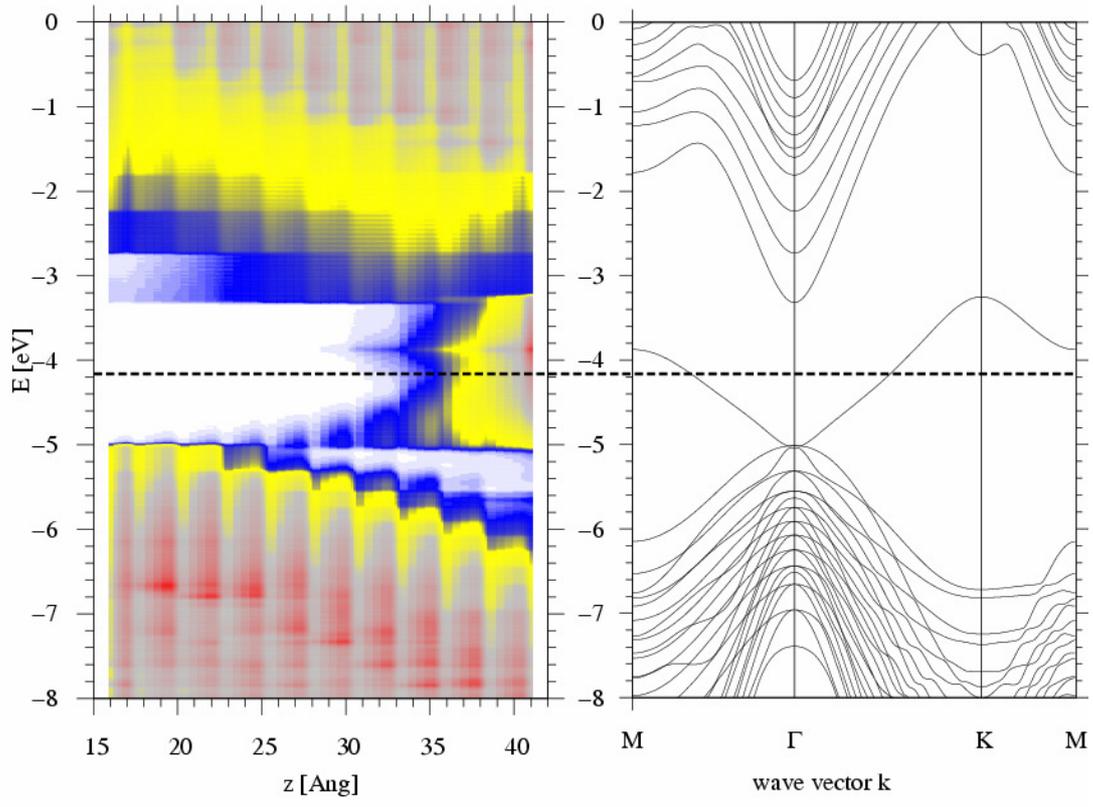

(a)

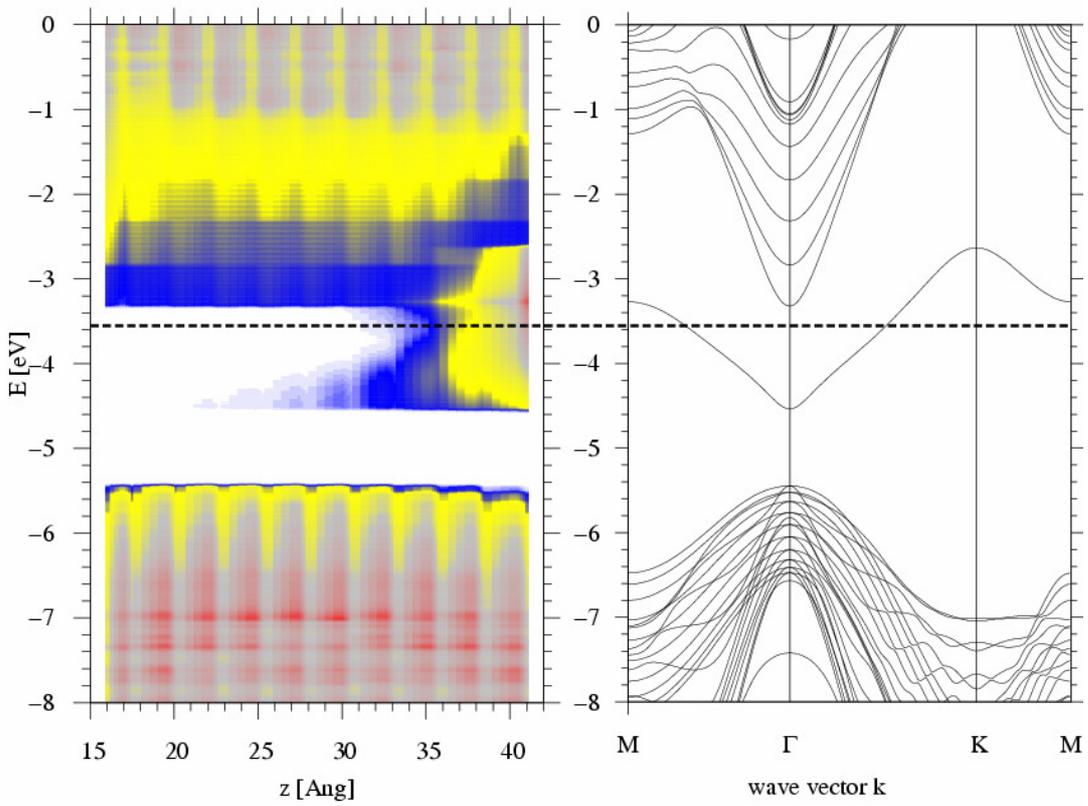

(b)

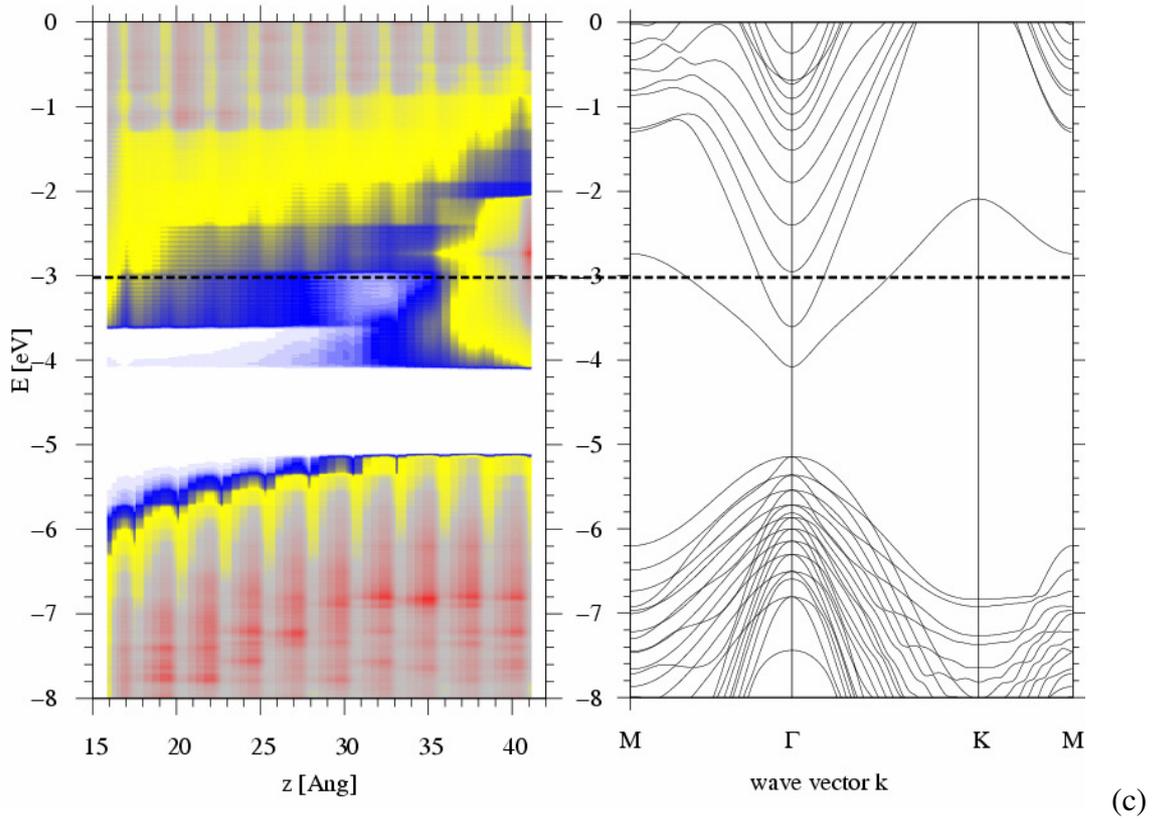

Fig. 3. Atom projected density of states (left) and the band diagram for the 10 DAL slab terminated by hydrogen atoms located at the 1.07 Å distance from nitrogen bottommost atoms: (a) Z = 0.70e, (b) Z = 0.73e and (c) Z= 0.78 e. The real GaN(0001) surface is positioned on the right side. Scale of the densities used is as follows: below 0.01- blue, 0.01 ÷ 0.1- yellow, 0.1 ÷1.0 – grey, above 1.0 – red.

These diagrams show spatial variation of the valence and conduction bands, of which the valence band is less diffuse, better suitable for analysis. In addition to the band states, the diagrams in Fig. 3 display considerable penetration of the surface states into the slab interior.

The Z = 0.73 e diagram shows the flatband case, the best suited for the determination of the gap. In this case Fermi level is located beneath the conduction band bottom, with no electric charge located on the band states. GaN(0001) surface state is partially filled and the Fermi level is pinned to the surface states.

Z = 0.70 e diagram shows also spatial linear dependence of the valence band states. The bottom of the surface states has approximately the same energy as the top of valence states on the other side of the slab. The band energy profile in Fig 3 and the potential profile in Fig.2 indicate that this case correspond to surface donor. Still the Fermi level is much higher than the VBM, therefore the slab interior is neutral and the potential profile is linear.

Finally, the diagram for Z = 0.78e is qualitatively different as it corresponds to the case of the degenerated electron gas. The Fermi level penetrates into the conduction band in the left (i.e. interior ) part of the slab, which causes occupation of portion of conduction band states that locates additional electric charge in the slab interior. This diagram presents nonlinear dependence of the valence band maximum in the left part of the slab, in accordance with the earlier presented potential profile. This affects the potential distribution and the valence band profile, changing this to parabolic i.e. strongly nonlinear. It is alos worth noting that the valence band energy is flat at the part close to the surface which is different from the potential profile which as presented in Fig. 2, changes across the whole thickness of the slab.

In order to investigate interplay of the surface and band states, a much wider slab of the same termination, consisting of 20 Ga-N DALs, was also simulated. The electron energy profiles (i.e. effectively inverted potentials) along the channeling position, both local and averaged over 3.3 Å, superimposed on the projected density of states, are plotted in Fig. 4. In order to compare the potential profiles with both valence and conduction bands, the two averaged profiles, shifted vertically by the DFT GaN bandgap, i.e. by 2.13 eV, are superimposed.

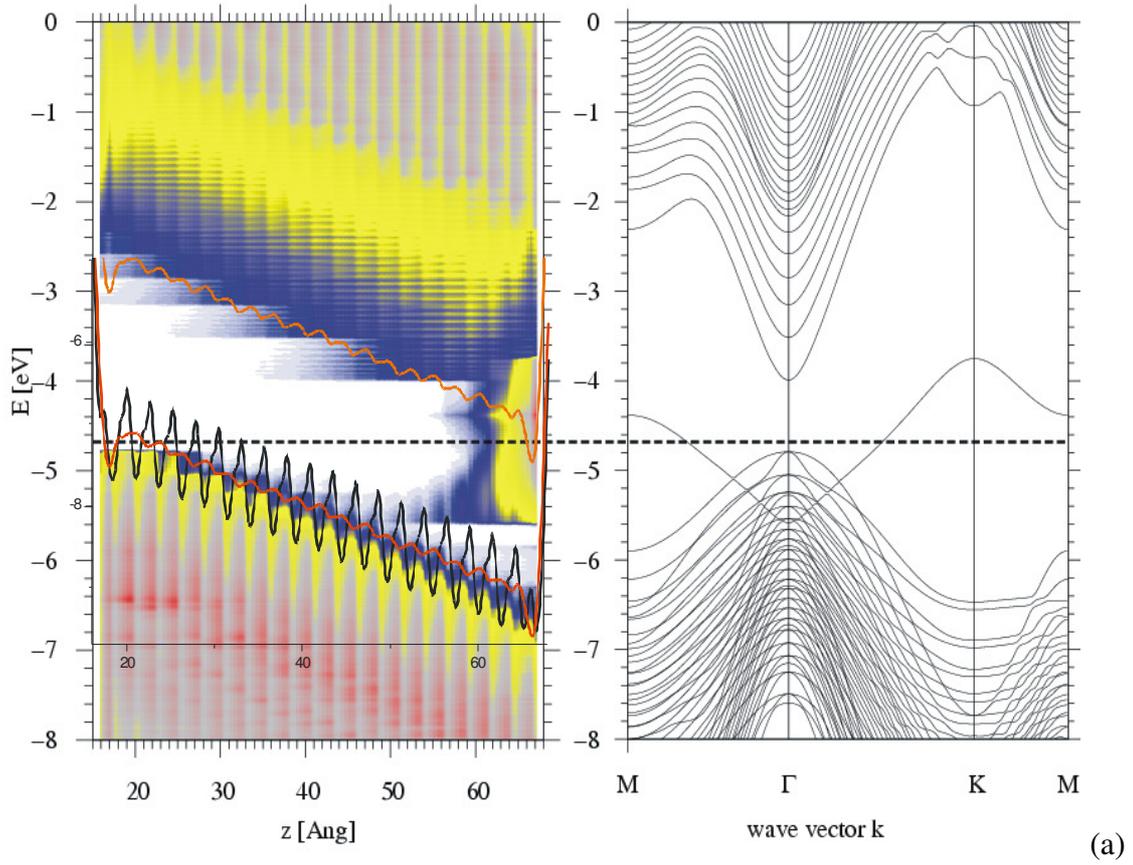

(a)

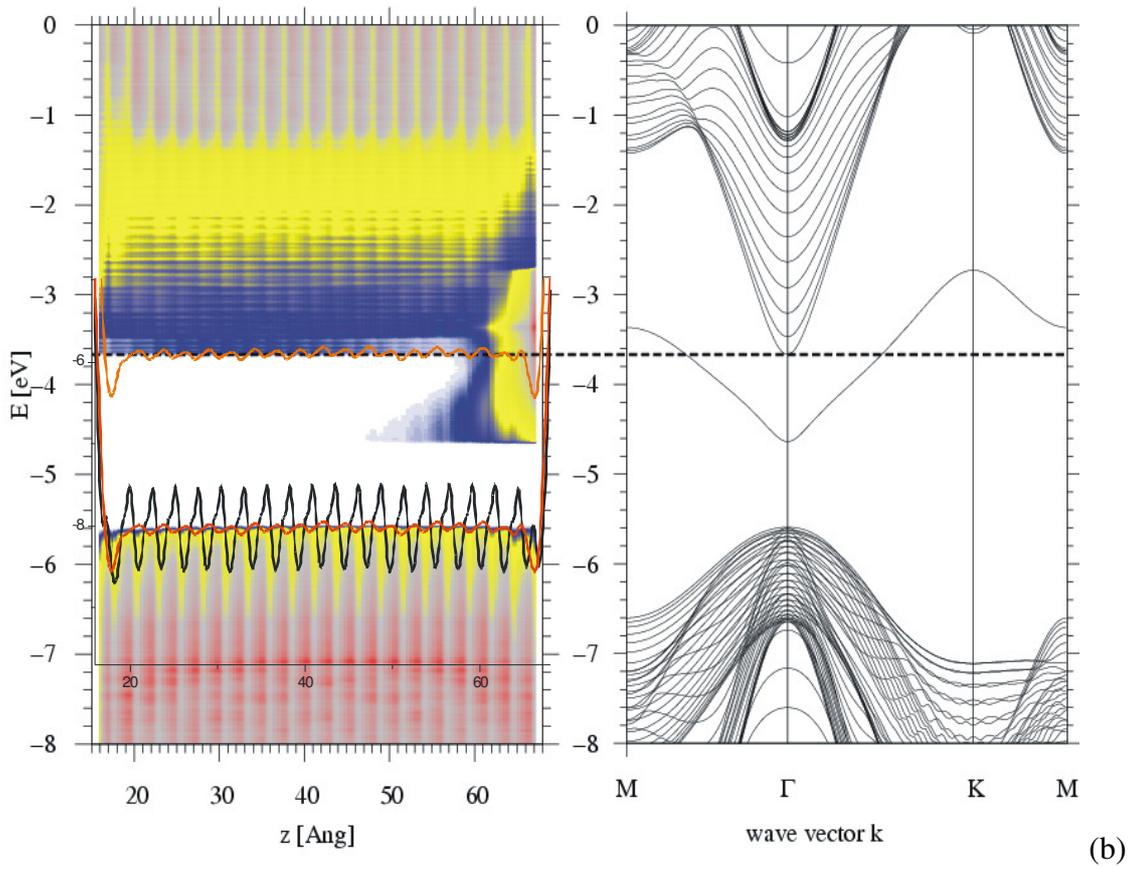

(b)

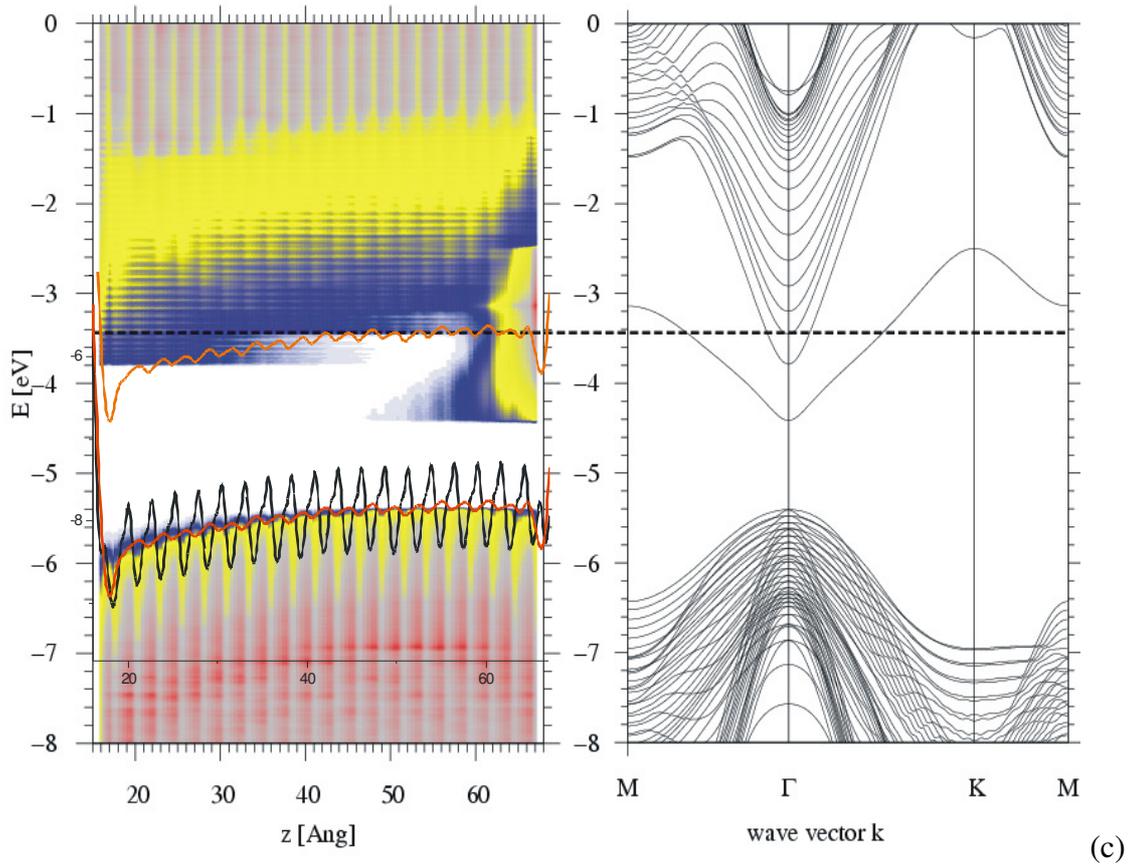

Fig. 4. Left diagrams - the profiles of electric potential, expressed as electron energy, in the slab consisting of 20 GaN DAL terminated by hydrogen atoms of the following fractional charge: (a) Z = 0.71 e, (b) Z = 0.735 e, (c) Z = 0.76 e . The black line denotes partially averaged (in x-y plane) potential profiles, adjusted to valence band maximum, the orange line denotes the profiles shifted by the GaN DFT energy gap equal to 2.13 eV, i.e. adjusted to conduction band minimum. The atom projected density of states are represented as follows: below 0.01 by blue, 0.01 ÷ 0.1 by yellow, 0.1 ÷1.0 by grey and above 1.0 by red color. Right diagrams present dispersion relations.

As it is shown in Fig.4, these electron energy profiles behave differently for the three above selection of hydrogen charges. The valence band change, linear over the entire slab, follows the potential profile exactly. The Fermi level is pinned by surface states at about 2 eV above valence band maximum. On the termination surface, the Fermi level is at the valence band maximum. Thus the projection of the spatially variable bands shifts the band by about 1.5 eV. In contrast to valence band, the spatial variation of the conduction band within the slab is different For most of the slab it is parallel to valence band, complying with the potential change but at the surface is shifted upward by about 1.5 eV..

The surface states presents considerable dispersion of more 2 eV in total, having an overlap with the conduction band states. Therefore, at the surface, the conduction band "repulsion" by the surface state is observed, so that the conduction band minimum is shifted up by more than 1.5 eV. The relative conduction band position at the surface is strongly shifted with respect to both valence band and surface state. The effect is purely quantum mechanical, as the long range electric potential is not affected.

In the two remaining cases, the long and short scale dependence is similar to the first. The valence band change follows closely the electrostatic potential. The long range variation of the conduction band is identical. As the Fermi level is pinned at about 1.5 eV above CBM, for the flat band case, the conduction band is shifted up by overlap with the surface state locally. The valence band is flat, unaffected by the surface state.

The case of the $Z = 0.76e$ hydrogen termination atoms displays different long range behavior of the electric potential. As the acceptor state energy is relatively close to conduction band, the Fermi level has to penetrate into the conduction band in the slab interior. Therefore the negative excess charge is accumulated there so that the potential profile is nonlinear. It is remarkable that the phenomenon of the conduction band repulsion by the surface state is visible again, changing the energy of the conduction band locally. Therefore this effect is universal, independent of the electric potential profile. As above, the valence band changes in accordance with the potential profile, showing no influence of the surface states.

Thus the relative positions of the bands and surface states are affected not only by long range electric field but the quantum overlap effects may change the energy of the band states locally, where substantial overlap of the wavefunction of the band and surface states occurs. In Ref 16, we have presented estimates of the relative motion of the surface states and the band states based on the data derived from the dispersion relations, as these shown in Fig.3. This approach could not reveal complex behavior of the states at the surface, shown in Fig. 3. Therefore the identification of the states energy were plagued by insufficient insight into the physics of the system and the conclusions described there were incorrect.

The change of the quantum states energy, is known as Stark effect, therefore the phenomenon was called Surface States Stark Effect (SSSE).[18] As discussed above, the physical nature of the SSSE stems from various contributions, long range electrostatic and local quantum mechanical. As expected, the field change of the quantum states energy depends on their location in space and also on their type, i.e. on their spatial extension. In order to Fig. 5 we plot the distribution of the electronic LDOS in the plane perpendicular to the surface is shown.

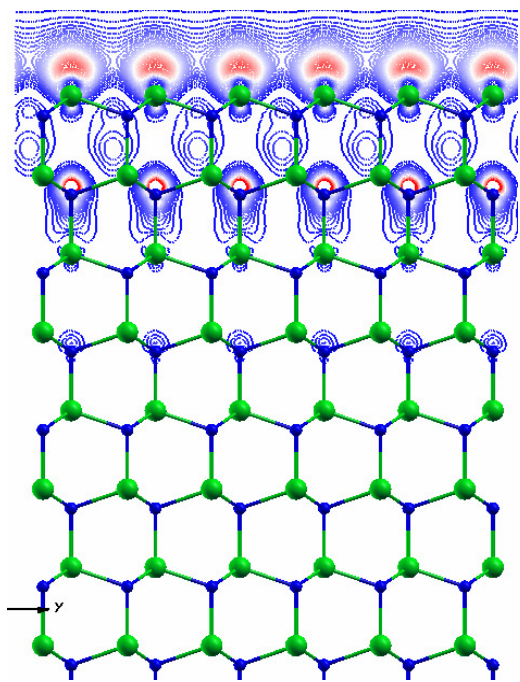

Fig. 5. Crossectional picture of the distribution of LDOS in the slab close to the GaN(0001) surface in the plane perpendicular to the surface.

As it is shown in Fig. 5, the distribution of the electronic charge is nonlocal, generating large overlap of the wavefunctions of the neighboring atoms, as confirmed by large dispersion of the surface states, plotted in Figs. 3 and 4. Therefore loading of the charge at the surface proceeds not via occupation of the separate states but by an increase of the average occupation of the band of surface states and gradual shift of the Fermi level. It is also worth noting that the surface states extend over two Ga-N DAL, i.e. relatively deep into the interior. Therefore the field at the surface affects their energy in the way similar to the band states.

As shown above, the average electric field may be deduced from the 10 DAL Ga-N slab results, as those shown in Fig. 4. Such functional dependence of the Fermi level, the energy of the bottom of the band of surface states, and the VBM and CBM on the electric field and the charge at the surface, obtained by gradual change of the charge of hydrogen termination atoms is shown in Fig. 6. The central part, (interval (1) – (5)) shows Fermi level located in the bandgap while the others are not. For higher charge of the termination atoms, in the interval (1)- (2), the Fermi level increases until at point (2) the Fermi level crosses the conduction band minimum. In this region the charge of the hydrogen is compensated by the negative band charge in the slab bulk, so the field is not affected. For very small charge of the hydrogen termination atoms, at the vicinity of the point (5), the Fermi level approximates the valence band, so the field does not change.

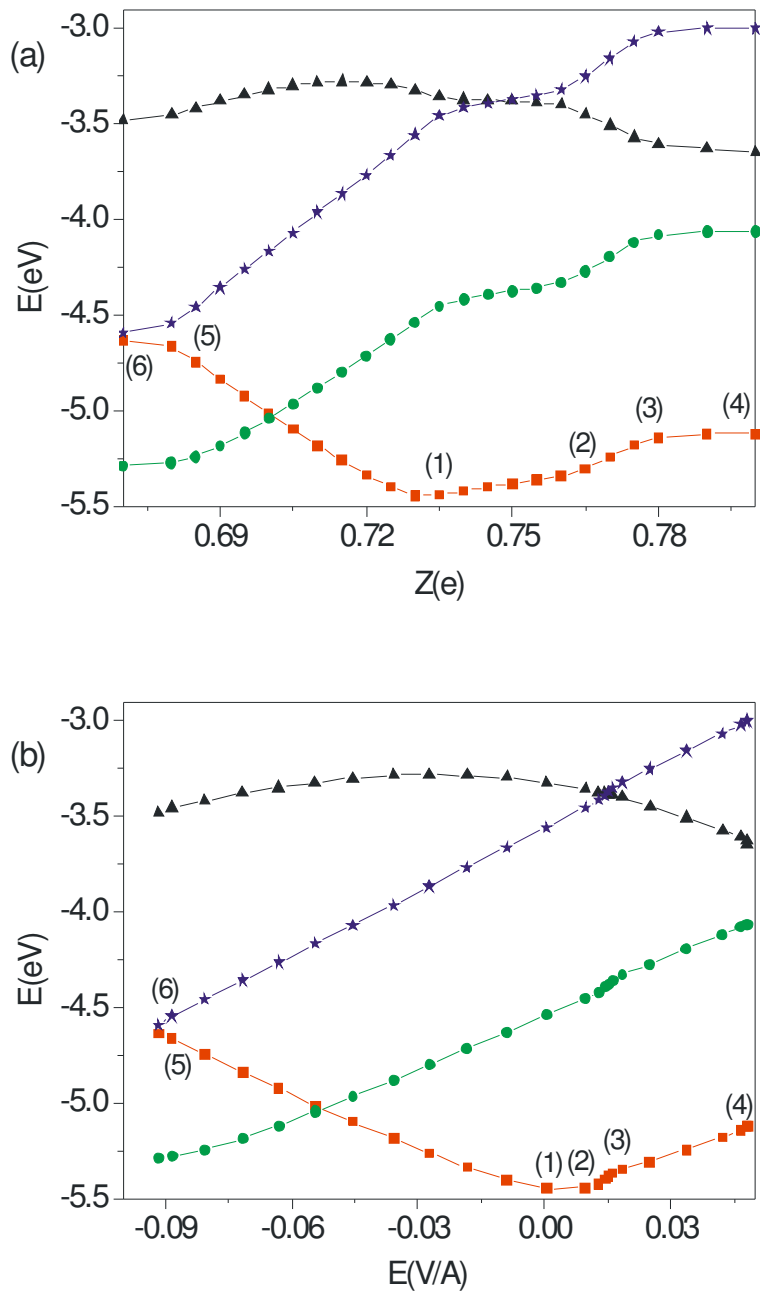

Fig. 6. VBM (red), CBM (black), Fermi energy (blue) and surface states minimum (SSM) (green) in function of the charge hydrogen termination atom (a) and the electric field at the surface (b). The electric field is defined as positive in the case when it is directed towards the surface, i.e. for the case of surface acceptor.

The DFT bandgap, obtained for zero field, is equal to:

$$E_g = E_c(0) - E_v(0) = 2.13 \text{eV} \tag{1}$$

therefore in order to recover the experimental bandgap, the SIESTA electron state energies should be scaled by the following factor:

$$\alpha = \frac{E_g(\exp)}{E_g(\text{DFT})} = \frac{3.4}{2.13} = 1.60 \tag{2}$$

The dependence of the Fermi energy on the electric field is

$$E_F(E) \cong -3.56 + 11.48 * E + 2.29 * E^2 \tag{3a}$$

that rescales to real values by multiplication by 1.6, to:

$$E_F(E) = -5.68 + 17.88 + 3.66 * E^2 \tag{3b}$$

where the field is expressed in V/Å. The nonlinear contribution is less than 0.2 of the linear variation, therefore Fermi energy is essentially linear function of the field at the surface. Note that the most of the variation is due to projection of the states due to the electric field as shown in Fig.1. The physically important is the difference between the Fermi energy and the energy of the bottom of the surface states as it reflects charging of the surface by gradual occupation of the band of surface states. The energy of the bottom of surface states is given by:

$$E_{sur}(E) \cong -4.60 + 9.66 * E + 13.04 * E^2 \tag{4a}$$

that rescales to real values as:

$$E_{sur} \cong -7.36 + 15.46 * E + 20.86 * E^2 \qquad (4b)$$

The difference, reflected slower variation of the bottom of the surface states, reflects Gauss theorem, i.e. the change of the field is by gradual occupation(charging) of the surface states. This difference, in real units, is expressed as.

$$\Delta E_{F-sur}(E) = E_F(E) - E_{sur}(E) \cong 1.68 + 2.42 * E - 17.20 * E^2 \qquad (5)$$

which for the simulated interval of the field values extending from -0.1 V/Å to 0.05 V/Å is dominated by the linear term. The total relative shift of these two energies over this interval is equal to 0.24 eV, and is therefore considerable. From the total the dominant part is for the acceptor i.e. for E > 0 is 0.16 eV while for donor part it is 0.08eV only. Note that the shift occurs in spite of the fact that Fermi level is pinned by the surface state, it is positive for E > 0, i.e. for surface acceptor as expected.

Another feature of considerable interest is the relative motion of the Fermi level and the conduction and valence bands, i.e. the band bending. As it was shown, the field related projection is responsible for most of the change of the band energies, obtained from dispersion relation or total density of states, obscuring relative motion of the valence band for E > 0 and the conduction band for E < 0. In addition, the conduction band bottom motion is strongly affected by the quantum overlap repulsion. Therefore the only reliable data can be derived from the valence band top motion for E < 0, for which the following relation was derived:

$$E_V(E) \cong -5.46 + 4.80*E + 49.93*E^2 \tag{6a}$$

after rescaling is:

$$E_V(E) \cong -8.74 + 7.68*E + 79.89*E^2 \tag{6b}$$

It is interesting to determine whether the relative position of the valence and surface states is affected by the field, which gives

$$\Delta E_{sur-V}(E) = E_{sur}(E) - E_V(E) \cong 1.36 + 7.78*E - 59.03*E^2 \tag{7}$$

It has to be concluded that the relative motion for E < 0 interval amounts to mere 0.24 eV. The relative motion of these energies, derived from dispersion relation in Ref 16, was due to the projection of the band energies. Additionally, using the above data the relative motion of the Fermi energy and the VBM is obtained:

$$\Delta E_{F-V} = E_F - E_V \cong 3.06 + 10.02*E - 76.21*E^2 \tag{8}$$

i.e. some motion of the Fermi level with respect to the band is observed. For the entire modeled region, i.e. from 0 to -0.05 V/A it amounts to 0.31 eV, thus slightly higher than the motion of the surface band which reflect charging of the surface states.

Finally, difference of the Fermi energy and the conduction band at the surface could defined in two ways, either by directly determined difference at the surface, i.e. accounting quantum repulsion of the surface and conduction band states or by the difference of the projected long distance dependence, which ignores local quantum repulsion effects. The latter is physically more relevant as it describes the field at the surface. This could be deduced from the difference between the bandgap, the energy of valence band maximum and the Fermi energy. Using Eq. 7, the Fermi energy difference with the long range projection of conduction band states is given as:

$$\Delta E_{C-F} = E_V + 3.4 eV - E_F \cong 0.34 - 10.02*E + 76.21*E^2 \quad (9)$$

Therefore at zero field, at clean GaN(0001) surface Fermi level is pinned 0.34 eV below the long distance projection of conduction band. The relative motion due to field effect reduces the difference for positively charged surface states, i.e. for surface donors by 0.31 eV, for the field equal to -0.05 V/Å.

It was shown recently however that the 2 x 1 reconstructed structure is slightly more stable than nonreconstructed 1 x 1 surface.[17] Therefore for comparison, 2 x 2 slab was simulated in order to find whether such reconstruction affects the electric properties of the surface. The results of the simulations are shown in Fig. 7.

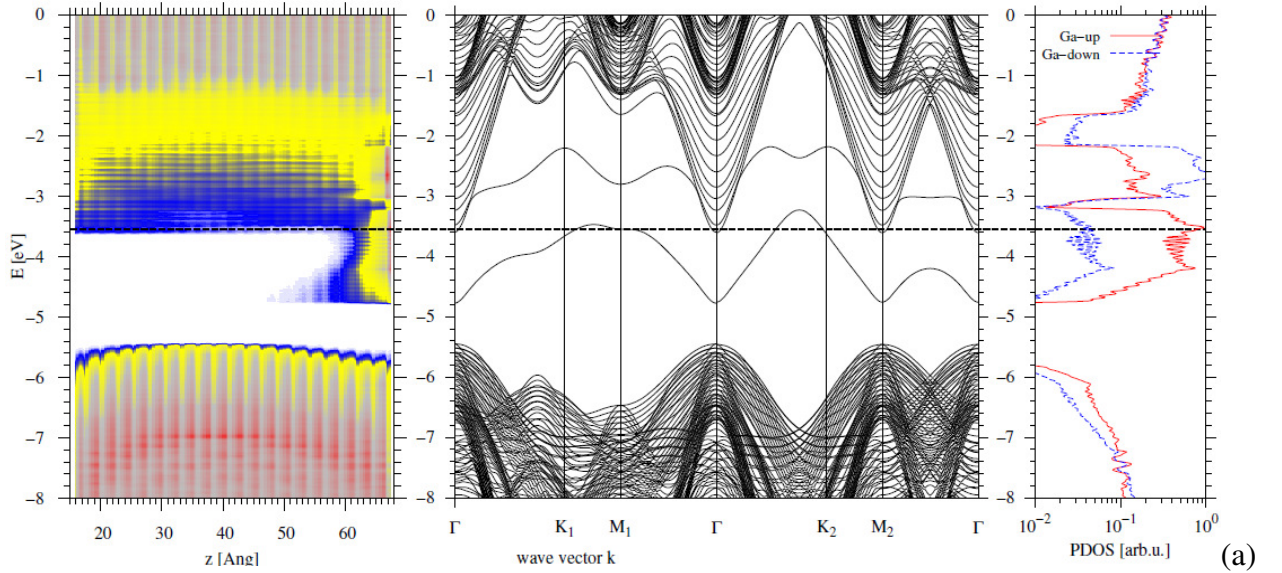

(a)

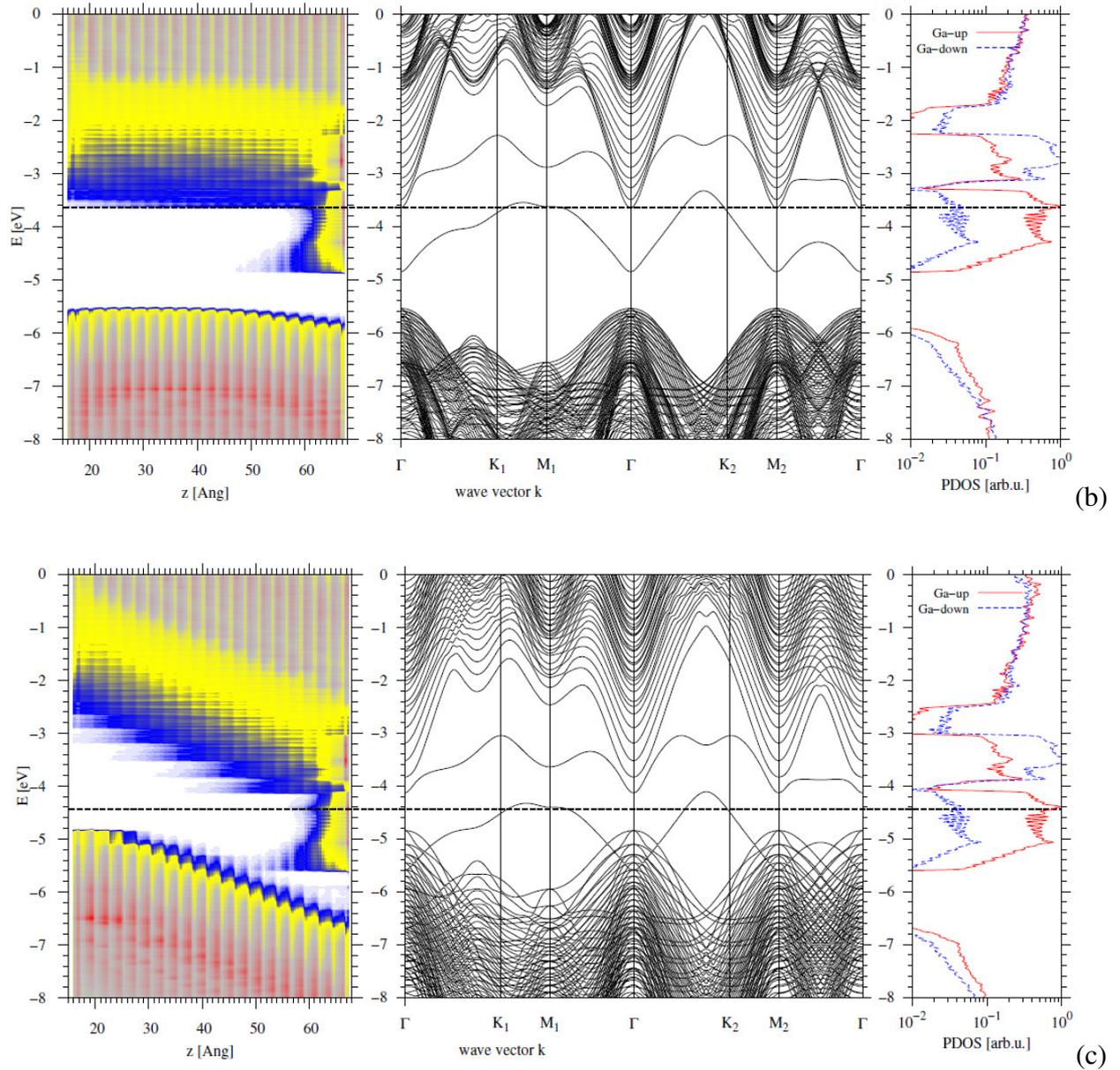

Fig. 7. The projected density of states (left), the band diagram and the total density of states of the gallium down (sp$^2$ hybridizied - blue line) and up (sp$^3$ hybridizied - red line) surface atoms (right), obtained for following fractional charges: (a) $Z = 0.76e_o$, (b) $Z = 0.735e_o$, (c) $Z = 0.71e_o$,. The real GaN(0001) surface is on the right side. Scale of the densities: below 0.01- blue, 0.01 ÷ 0.1- yellow, 0.1 ÷1.0 – grey, above 1.0 – red.

These data are compatible with the earlier findings, the surface state band is split into two subbands, of which the lower energy is totally occupied, and the higher is empty. The Fermi level is positioned at the top of lower subband. Altogether, the energy of band states and the Fermi level are compatible with the results obtained for plain surface.

## IV. Summary.

Simulation of semiconductor surfaces by slab models leads to emergence of the electric field within the slab that is controlled by the charge distribution in the surface states, the termination surface states and possible by the band states. The latter case is realized when the Fermi level penetrates into the band in the slab interior. The field shifts the energy of the surface states, i.e. it leads to emergence of Surface States Start Effect (SSSE). Detailed investigation of the nature of the surface state motion revealed complicated scenario realized at GaN(0001) surface. It is shown that the two basic scales describe electric properties of the subsurface layer. The long scale change of the electric potential is induced by the charge of the surface. The valence band follows this long scale change of the electric potential. The long distance change of the energy of the conduction band is similar but its short distance energy change is controlled by quantum overlap repulsion of the surface and the conduction band states.

In the case of slab modeling the energy of the states is shifted by the electric field leading to variation of the states energy depending on their location within the slab. This leads to projection of the states and artificial narrowing of the bandgap. The projection effect is responsible for majority of the energy shift observed in the dispersion relation in Ref 16.

The states at GaN(0001) surface are extended both in the direction perpendicular to the surface, and also in the surface plane. Their in plane extension leads to large overlap with the neighboring atoms and creation of the band of the surface states having dispersion above 1.5 eV. The surface states extend into the interior to about 4 Ga-N DALs. Thus they are similar to the band states, therefore they react similarly to the applied electric field. The relative change of the energy of the band and surface states is only 0.24 eV for the electric field changing from 0 to -0.05 V/Å, thus a small fraction of that reported in Ref 16.

The accumulation of the charge at the surface proceeds via gradual change of the occupation of the surface band. The Fermi level is pinned about 0.34 eV below the long distance projection of the conduction band bottom. The motion due to occupation of surface states, moves the Fermi level by 0.31 eV for the electric field up to 0.05 V/Å.

The 2 x 1 reconstruction, that is energy stable surface structure of clean GaN(0001) surface does not bring any significant change of its electric properties. The surface band is divided into two subbands of which the upper, due to $p_z$ states is empty and lower is occupied. The Fermi energy remains at approximately the same position as for the nonreconstructed surface.

## Acknowledgements

The calculations reported in this paper were performed using computing facilities of the Interdisciplinary Centre for Modelling (ICM) of Warsaw University. The research published in this paper was supported by Poland's Ministry of Science and Higher Education grant no. POIG 01.01.02-00-008/08.